\documentclass[pra,superscriptaddress,twocolumn,showpacs,balancelastpage]{revtex4}
\usepackage{amsfonts}
\usepackage{amsmath}
\usepackage{amssymb}
\usepackage{graphicx}

\begin{document}

\title{Dynamical regimes of a quantum SWAP gate beyond the Fermi golden rule.%
}
\author{Axel D. Dente }
\email[]{dente@famaf.unc.edu.ar}
\affiliation{Facultad de Matem\'{a}tica, Astronom\'{\i}a y F\'{\i}sica, and
Instituto de F\'{\i}sica (CONICET), Universidad Nacional de C\'{o}rdoba,
Ciudad Universitaria, 5000 C\'{o}rdoba, Argentina.}

\author{Ra\'{u}l A. Bustos-Mar\'{u}n }
\affiliation{Facultad de Matem\'{a}tica, Astronom\'{\i}a y F\'{\i}sica, and
Instituto de F\'{\i}sica (CONICET), Universidad Nacional de C\'{o}rdoba,
Ciudad Universitaria, 5000 C\'{o}rdoba, Argentina.}
\affiliation{Departamento de Fisicoqu\'{\i}mica. Facultad de Ciencias Qu\'{\i}%
micas, Universidad Nacional de C\'{o}rdoba, Ciudad Universitaria, 5000 C\'{o}%
rdoba, Argentina.}
\author{Horacio M. Pastawski }
\affiliation{Facultad de Matem\'{a}tica, Astronom\'{\i}a y F\'{\i}sica, and
Instituto de F\'{\i}sica (CONICET), Universidad Nacional de C\'{o}rdoba,
Ciudad Universitaria, 5000 C\'{o}rdoba, Argentina.}

\begin{abstract}
We discuss how the bath's memory affects the dynamics of a swap gate. We
present an exactly solvable model that shows various dynamical transitions
when treated beyond the Fermi Golden Rule. By moving continuously a single
parameter, the unperturbed Rabi frequency, we sweep through different
analytic properties of the density of states: (I) collapsed resonances that
split at an exceptional point in (II) two resolved resonances ; (III)
out-of-band resonances; (IV) virtual states; and (V) pure point spectrum. We
associate them with distinctive dynamical regimes: overdamped, damped
oscillations, environment controlled quantum diffusion, anomalous diffusion
and localized dynamics respectively. The frequency of the swap gate depends
differently on the unperturbed Rabi frequency. In region I there is no
oscillation at all, while in the regions III and IV the oscillation
frequency is particularly stable because it is determined by the
environment's band width. The anomalous diffusion could be used as a
signature for the presence of the elusive virtual states.
\end{abstract}
\received{6 October 2008}

\maketitle

\section{Introduction}

It is well known that processing information using quantum mechanics makes
possible communication procedures and computational tasks that could
outperform classical devices in terms of security or speed \cite%
{Bennet-DiVincenzo}. There are extensive experimental efforts in many fields
to realize implementations of the necessary building blocks for quantum
information processing. The key ingredient is an externally controlled
evolution of a superposition state which implies a form of quantum
parallelism. The alternatives range from electrons spins \cite{vanderSypen},
superconducting circuits \cite{Urbina, Nakamura}, quantum electrodynamic
cavities \cite{Wallraff, Haroche}, and optical ion traps \cite{Blatt}, to
nuclear magnetic resonance (NMR) in liquids \cite{Cory-Review} and solid
state \cite{ADLP06}. One of the most important building blocks is the swap
gate, where a system $\mathcal{S}$ jumps between two degenerate states, $A$
and $B$, when the coupling $V_{AB}$ is turned on. Starting on state $A$, the
return probability oscillates with the Rabi frequency $\omega
_{0}=2V_{AB}/\hslash $. However, it is clear that it is not feasible to
isolate completely any real system. In practice, the interactions with an
environment $\mathcal{E}$ \cite{Zurek2003, Myatt-Nature} perturb the
evolution, smoothly degrading the quantum interferences with a
\textquotedblleft decoherence\textquotedblright\ rate, $1/\tau _{\phi }$.
This rate is usually identified with the $\mathcal{SE}$ interaction rate $%
1/\tau _{\mathcal{SE}}$, typically evaluated from the Fermi Golden Rule
(FGR). In spin systems, $A$ and $B$ could be the $\uparrow \downarrow $ and $%
\downarrow \uparrow ~$spin configurations respectively. Weak interactions ($%
1/\tau _{\mathcal{SE}}^{{}}\ll 2\omega _{0}$) produce a slightly slower
oscillation which decays at a rate $1/\tau _{\phi }=1/(2\tau _{\mathcal{SE}%
}^{{}})$. There are experimental conditions, however, where the observed
frequency\ shows a dynamical transition on its dependence of the $\mathcal{SE%
}$ interaction \cite{ADLP06}. In fact, the swapping frequency is a\textit{\
non-analytic} function of the interaction rate. At a critical strength $%
1/\tau _{\mathcal{SE}}^{c}=2\omega _{0}$, the oscillation freezes indicating
a \textit{transition} to a new\textit{\ dynamical regime}. The initial state
now decays to equilibrium at a slower rate $1/\tau _{\phi }^{{}}\propto
\omega _{0}^{2}~\tau _{\mathcal{SE}}^{{}}$, which vanishes for strong $%
\mathcal{SE}$ interaction. This last regime can be seen as a quantum Zeno
phase, where the dynamics is inhibited by the frequent \textquotedblleft
observations\textquotedblright\ \cite{Misra-Sudarshan} of the environment.
Such quantum freeze can arise as a pure dynamical process, governed by
strictly unitary evolutions (see Refs. \cite{Pascazio, Pastawski-Usaj}).
Indeed, some of the phenomenology of that transition was not foreign to
spectroscopists. The collapse of the independent resonance lines leads to
the exchange and motional narrowing addressed by Van Vleck \cite{van Vleck}
and Bloembergen, Purcell, and Pound \cite{Bloembergen} in the 1940's and
synthesized in the analytic properties of a phenomenological classical
probabilistic model by P. W. Anderson \cite{Anderson}. The quantum
description of the phase transition \cite{ADLP06} required a self-consistent
calculation of the oscillation in presence of the $\mathcal{SE}$
interaction. While the system remembers its previous state, the environment
was described in the \textquotedblleft fast fluctuations\textquotedblright\
approximation where it retains no memory of its previous state \cite%
{Alvarez-PRA}.

In this work we present a simple and exactly solvable quantum model that,
while only describing the coherent part of a spin SWAP gate dynamics \cite%
{Danieli-ChPhLett}, presents the dynamical transitions. Our goal is to
deepen the understanding of how bath's memory affects the system dynamics
beyond the Fermi golden rule. A complementary vision for the dynamics under
the action of a given Hamiltonian is, of course, the spectral representation
which can be studied as function of the $\mathcal{S-E}$ interaction strength.
However, the energy representation hides much useful dynamical information
in subtle spectral properties, such as resonances that collapse at the
"exceptional points" (EPs) in the complex plane. Other unusual properties
involve resonances that shrink and jump into the nonphysical Riemann sheet
to become virtual states, which sometimes are also called antibound states%
\cite{Taylor, Nussenzveig}. Ultimately, these resonances can transform
themselves into isolated singularities on the real axis, accounting for
localized\ states. In our model all of these transitions will appear
naturally through the variation of a \textit{single} control parameter.
There are a number of physical systems that show some of these delicate
spectral properties: The EP or collapse of resonances has been observed in
crystals of light \cite{Zeilinger}, electronic circuits \cite{Scholtz},
propagation of light in dissipative media \cite{Shuvalov, Berry}, vacuum
Rabi splitting in semiconductors cavities \cite{Gibbs}, in microwave
billiards \cite{RichterPRL01, Richter 2007, RichterPRE04, RichterPRL03},
detuned but coupled piano strings \cite{Weinreich}, and there are a number
of examples drawn from electron-paramagnetic resonance \cite{Calvo}, and
solid state\ NMR \cite{ADLP06}. It also may appear in many theoretical
models: e.g., describing the decay of superdeformed nuclei \cite%
{BarrettPRC99}, phase transitions and avoided level crossings \cite%
{HeissPRA91, Mossmann, HeissPRE00}, coupling of bound states to open decay channels \cite{Rotter95}, geomagnetic polarity reversal \cite%
{StefaniPRL05}, tunneling between quantum dots \cite{BarrettPSSB02,
DanieliSSC}, optical microcavity \cite{LonghiPRA06}, vibrational surface
modes \cite{CP06}, and in the context of the crossing of two Coulomb
blockade resonances \cite{Weidenmuller}. On the other hand, the
virtual-localized transition has been addressed in the context of the famous
$n-p$ singlet system \cite{Taylor}, models of stabilization of quantum
mechanical binding by potential barriers \cite{Hog95}, Feshbach resonances
\cite{Marcelis, Pupasov} and stability of atomic and molecular states \cite%
{serra-Moiseyev-stable,Yamashita}.

This paper is organized as follows. In Sec. II the model is presented. It
is solved in Sec. III using a Green%
%TCIMACRO{\U{b4}}%
%BeginExpansion
\'{}%
%EndExpansion
s function formalism \cite{Pasta-Medina} and its analytical properties are
analyzed. In Sec. IV the different parametric regions are associated with
the analytic properties of the density of states: region I) collapsed
resonances; region II) resolved resonances; region III) out of band
resonances; region IV) virtual states; and region V) pure point spectrum.
These parametric regions are associated with distinctive dynamical regimes:
overdamped decay, damped oscillations, environment controlled quantum
diffusion, anomalous diffusion and localized dynamics, respectively. The
anomalous diffusion then could be considered as a signature of virtual
states. The summarizing conclusions are presented in the last section.

\section{The Model}

\begin{figure}[ptb]
\begin{center}
\includegraphics[height=0.7572in,width=3.066in]
{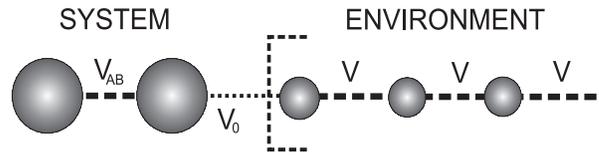}
\end{center}
\caption{Representation of the system and the environment. The first two
sites are connected through $V_{AB}$. The hopping amplitude inside the
environment is $V$. The system-environment interaction is $V_{0}$.}
\label{Sistema}
\end{figure}

Our model, schematized in Fig. \ref{Sistema}, is a tight-binding
semi-infinite linear chain. The Hamiltonian has one part that describes the
two states of the system, another that describes the infinite degrees of
freedom of an environment and a third part that is the interaction between
both. It is expressed as%
\begin{equation}
H=H_{\mathcal{S}}+H_{\mathcal{E}}+V_{\mathcal{S-E}},  \label{Hamiltoniano}
\end{equation}%
where%
\begin{equation}
H_{\mathcal{S}}=E_{A}\left\vert A\right\rangle \left\langle A\right\vert
+E_{B}\left\vert B\right\rangle \left\langle B\right\vert -V_{AB}\left(
\left\vert A\right\rangle \left\langle B\right\vert +\left\vert
B\right\rangle \left\langle A\right\vert \right)
\end{equation}%
\begin{equation}
H_{\mathcal{E}}=\sum\limits_{n=1}^{\infty }E_{n}\left\vert n\right\rangle
\left\langle n\right\vert -V\left( \left\vert n\right\rangle \left\langle
n+1\right\vert +\left\vert n+1\right\rangle \left\langle n\right\vert
\right)
\end{equation}%
\begin{equation}
V_{\mathcal{S-E}}=-V_{0}\left( \left\vert B\right\rangle \left\langle
1\right\vert +\left\vert 1\right\rangle \left\langle B\right\vert \right) ,
\end{equation}%
and $\left\vert n\right\rangle $, with $n\geq 1$, is the state localized at
the $n$th site of the chain. $V_{AB}$, $V_{0}$ and $V$ are the positive
hopping amplitudes between two contiguous sites. The first two sites, linked
by $V_{AB}$, are what we call \textit{the system }$\mathcal{S}$. The rest of
the chain is what we take as \textit{the environment }$\mathcal{E}$. In this
work, we are interested in the case $E_{A}=E_{B}=E_{n}=0$. This choice of
parameters together with the asymmetric connection to the environment
enables a rich spectrum.

%EndExpansion

This model applies to implementations of NMR quantum information processors
by resorting to the Jordan-Wigner transformation \cite{ADLP06}\cite{DPL04}
that maps spins into a Fermionic system. More specifically, spin chains with
XY interaction are transformed, at the experimentally relevant high
temperatures, into a system of independent fermions in a tight binding chain
\cite{Madi-Ernst-ChPhLett1997}. In particular, Ref. \cite{ADLP06} shows how
the Swap Gate in solid state NMR, is obtained from a coherent dynamics as
that of our model plus an incoherent contribution. The fermionic model,
having incorporated the temperature in the appropriate regime, is a natural
alternative to the various spin-boson models, where a number of transitions
are also known to exist as a function of temperature and interaction
strength \cite{spin-boson}\textbf{.}

\section{Analytic Solution}

In this work we use the Green's Function (GF) formalism to solve the model.
Starting with the isolated system $H_{\mathcal{S}}$, we define the GF matrix
$\mathbb{G}$,%
\begin{equation}
\mathbb{G}^{(0)}\left( \varepsilon \right) =\left( \mathbb{H}_{\mathcal{S}%
}-\varepsilon \mathbb{I}\right) ^{-1}.
\end{equation}%
In order to make a fully consistent definition of $G_{n,m}(\varepsilon )$ as
the Fourier transform of the retarded propagator $G_{n,m}(t)$ we assume
that each site $n$ has an intrinsic decay process $E_{n}\rightarrow E_{n}-%
\mathrm{i}\eta _{n}$. This solves the Gutzwiller objection on this aspect
\cite{Gutzwiller}$.$ However, since these imaginary parts are considered
infinitesimal, they are not written explicitly in what follows.

The first diagonal component of $\mathbb{G}\left( \varepsilon \right) $
gives us information about the system dynamics. This component becomes%
\begin{equation}
G_{AA}^{(0)}\left( \varepsilon \right) =\frac{1}{\varepsilon -\dfrac{%
V_{AB}^{2}}{\varepsilon }}.  \label{Green de 2 sitios}
\end{equation}%
Here, we can see that the poles of $G_{AA}^{(0)}\left( \varepsilon \right) $
give the system bonding and antibonding energies $E_{1,2}=\pm V_{AB}$.
We
consider an isolated system of two sites and excite the first one $%
\left\vert 1,0\right\rangle \equiv \left\vert A\right\rangle $. The system
evolves oscillating between the states $\left\vert A\right\rangle $ and $%
\left\vert 0,1\right\rangle \equiv \left\vert B\right\rangle $ with a
characteristic Rabi frequency $\omega _{0}=\frac{2V_{AB}}{\hbar }$. This
oscillation is used to generate a SWAP gate by letting the Hamiltonian act
during a time $\ t_{\mathrm{swap}}=\frac{5\pi }{2}\frac{1}{\omega _{0}}$.

When we take into account the environment, the first diagonal component of
the GF becomes
\begin{equation}
G_{AA}\left( \varepsilon \right) =\frac{1}{\varepsilon -\dfrac{V_{AB}^{2}}{%
\varepsilon -\dfrac{V_{0}^{2}}{V^{2}}\Sigma \left( \varepsilon \right) }}.
\label{Green de inf Sitios}
\end{equation}%
Typically the self-energy $\Sigma \simeq \Delta -\mathrm{i}\Gamma $ is
evaluated within a Fermi golden rule approximation \cite{PascazioFGRbook} as
an $\varepsilon $ independent complex number. Since this would imply
neglecting all dynamics and memory effects of the environment, such
procedure could miss some subtle behaviors \cite%
{Khalfin,Ghirardi,GarciaCaderon-Moshinsky}. Our model enables the evaluation
of the exact self-energy of an environment represented by the semi-infinite
chain, and hence accounts precisely for these \textquotedblleft memory
effects\textquotedblright . According to the continued fractions solution
\cite{Pasta-Medina} of the renormalized perturbation expansion \cite%
{Economou},%
\begin{align}
\Sigma\left(  \varepsilon\right)   &  =\frac{V^{2}}{\varepsilon-\dfrac{V^{2}%
}{\varepsilon-\dfrac{V^{2}}{\varepsilon-%
%TCIMACRO{\QDATOP{{}}{\ddots}}%
%BeginExpansion
\genfrac{}{}{0pt}{0}{{}}{\ddots}%
%EndExpansion
}}}\label{Dyson-chain-general}\\
&  =\frac{V^{2}}{\varepsilon-\Sigma\left(  \varepsilon\right)  }.
\label{Dyson-chain-infinite}%
\end{align}%
which sums up to the form%
\begin{equation}
\Sigma \left( \varepsilon \right) =\Delta \left( \varepsilon \right) -%
\mathrm{i}\Gamma \left( \varepsilon \right) ,
\end{equation}%
with%
\begin{equation}
\Delta \left( \varepsilon \right) =\left\{
\begin{array}{cc}
\frac{\varepsilon }{2}-\sqrt{\left( \frac{\varepsilon }{2}\right) ^{2}-V^{2}}
& \varepsilon >2V \\
\frac{\varepsilon }{2} & \left\vert \varepsilon \right\vert \leq 2V \\
\frac{\varepsilon }{2}+\sqrt{\left( \frac{\varepsilon }{2}\right) ^{2}-V^{2}}
& \varepsilon <-2V,%
\end{array}%
\right.   \label{Delta-E}
\end{equation}%
and%
\begin{equation}
\Gamma \left( \varepsilon \right) =\left\{
\begin{array}{cc}
0 & \varepsilon >2V \\
\sqrt{V^{2}-\left( \frac{\varepsilon }{2}\right) ^{2}} & \left\vert
\varepsilon \right\vert \leq 2V \\
0 & \varepsilon <-2V.%
\end{array}%
\right.   \label{Gamma-E}
\end{equation}%
A brief commentary on the complex self-energies, i.e., non-Hermitian terms,
is in order. Its appearance, either in a FGR calculation or in the exact
solution of Eq. \ref{Dyson-chain-infinite}, relays on the fact that the new
eigenstates are completely orthogonal to the unperturbed ones. In our case
they are extended states enabled by the consideration of the thermodynamic
limit: the number of states becomes infinite  before $\eta $ is allowed to
reach 0 \cite{Horacio-FGR}. As discussed above, adding this small imaginary
part means putting the system in contact with an additional environment $%
\mathcal{E}^{\prime }$. In this situation, Eq. \ref{Dyson-chain-infinite}
produces a self-energy containing a square root function of the energy
instead of a ratio among polynomials that results when Eq. \ref%
{Dyson-chain-general} is applied to a finite system. The sign in front of
the square root in Eqs. \ref{Delta-E} and \ref{Gamma-E} is chosen to ensure
the physical (i.e. decaying) behavior when $\eta <0$. As discussed in Ref.
\cite{Horacio-FGR} the presence of non-Hermitian terms is fundamental in
allowing a dynamical phase transition. In the general context of
non-Hermitian quantum mechanics a similar conclusion holds \cite{Pablo-EPL}.

The poles of Eq. \ref{Green de inf Sitios}, control the dynamics of the
system and can be obtained analytically as

\begin{equation}
\varepsilon _{r}^{2}=\frac{V_{AB}^{2}\left( 2V^{2}-V_{0}^{2}\right)
-V_{0}^{4}\pm V_{0}^{2}\sqrt{\left( V_{AB}^{2}+V_{0}^{2}\right)
^{2}-4V_{AB}^{2}V^{2}}}{2\left( V^{2}-V_{0}^{2}\right) }.  \label{Polos}
\end{equation}%
\begin{figure}[tbp]
\begin{center}
\includegraphics[height=4.1148in,width=3.1687in]
{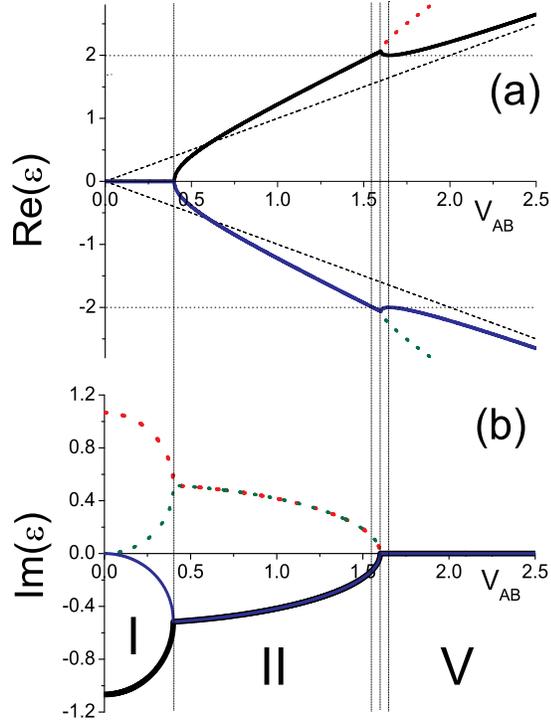}
\end{center}
\caption{(Color online) Real (a) and imaginary (b) part of the Green
function poles vs. $V_{AB}$ for $V_{0}=0.8$. All energies and hoppings are
in $V$ units. (a) The dashed line represent the poles for an isolated system.
Different colors identify the poles. The nonphysical ones are represented
with dotted lines. The vertical dotted lines divide the dynamical regions.}
\label{Re + Im-Polos}
\end{figure}
\begin{figure}[tbp]
\begin{center}
\includegraphics[height=4.1079in,width=3.0692in]
{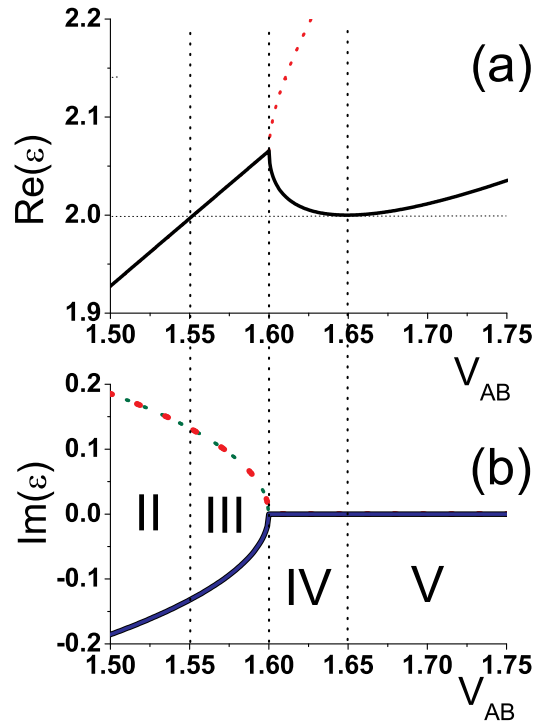}
\end{center}
\caption{(Color online) Zoom of Fig. \protect\ref{Re + Im-Polos} in the
region where virtual states and out of band resonances occur. All energies
and hoppings are in $V$ units.}
\label{Re-Im chico}
\end{figure}
Equation \ref{Polos} has four solutions, see Fig. \ref{Re + Im-Polos}. When they
have imaginary parts, only those with negative ones represent a decaying
response to an initial condition. These imaginary parts are precisely the
exponential decay rate in the self-consistent Fermi golden rule \cite{FP06}.
When the four poles are real, the physical ones should approach to the
isolated system poles (shown in Fig. \ref{Re + Im-Polos} with dashed lines).
\ The real and imaginary parts of the poles are shown as a function of the
system hopping $V_{AB}$ for fixed values of $V_{0}=0.8V$. The solutions not
satisfying the above conditions are indicated with dotted lines. Hereafter,
we will refer as \textquotedblleft the Poles\textquotedblright\ of the GF
only\ to those indicated by the continuous line.

The Local Density of States (LDOS), can be evaluated as%
\begin{equation}
N_{A}\left(  \varepsilon\right)  =-\frac{1}{\pi}\lim_{\eta\rightarrow0^{+}%
}\operatorname{Im}G_{AA}\left(  \left\{  E_{n}-\mathrm{i}\eta_{n}\right\}
_{\forall n},\varepsilon\right)  .
\end{equation}
This definition is equivalent to the standard one, $-\frac{1}{\pi}\lim
_{\eta\rightarrow0^{+}}\operatorname{Im}G_{AA}$($\varepsilon+\mathrm{i}\eta$), for most practical purposes.

An advantageous feature of the present model is that the LDOS for $%
\left\vert \varepsilon \right\vert \leq 2V$ can be factorized as
\begin{equation}
N_{A}\left( \varepsilon \right) =N_{1}\left( \varepsilon \right) \times
L_{1}\left( \varepsilon \right) \times L_{2}\left( \varepsilon \right) .
\label{LDoS Overdamped Equation}
\end{equation}%
Here, $L_{1}$ and $L_{2}$, are Lorentzian functions (LFs), and $N_{1},$is
the \textit{density of directly connected states} (i.e. the LDOS of the
first site of the semi infinite chain).%
\begin{equation}
N_{1}\left( \varepsilon \right) =\frac{1}{\pi V^{2}}\sqrt{V^{2}-\frac{%
\varepsilon ^{2}}{4}}.
\end{equation}%
The LFs\ $L_{1}\left( \varepsilon \right) \ $and $L_{2}\left( \varepsilon
\right) $ are related with the real and the imaginary part of the GF's
poles. Their centers move with the real part of the poles, and their widths
are determined by the imaginary parts.

In Fig. \ref{Re + Im-Polos} it is observed that there are regions with
different analytical behaviors, some of them separated by abrupt changes
that are consequence of the nonanalytical behavior of the GF poles. The
difference between the physical poles has a real part $\widetilde{\omega
}=\left\vert \operatorname{Re}\left(  \varepsilon_{r1}\right)
-\operatorname{Re}\left(  \varepsilon_{r2}\right)  \right\vert =2\left\vert
\operatorname{Re}\left(  \varepsilon_{r1}\right)  \right\vert $, representing an effective Rabi
frequency. The imaginary part is associated with the decay rate into
environment's states. This work mainly focuses on the study of the behavior
of $\widetilde{\omega }$, as the parameter that characterizes the dynamics.

The relation between GF's analytic properties and the dynamics is clarified
by writing the survival probability in the energy-time representation
%\begin{widetext}
\begin{eqnarray}
%\begin{align}
P_{AA}\left(  t\right)   &  =%
{\textstyle\int\limits_{-\infty}^{\infty}}
\mathrm{d}\varepsilon%
{\textstyle\int\limits_{-\infty}^{\infty}}
\frac{\mathrm{d}\omega}{2\pi}G_{AA}(\varepsilon+\tfrac{1}{2}\hbar\omega) \nonumber \\
& {\times} G_{AA}^{\ast}(\varepsilon-\tfrac{1}{2}\hbar\omega)\exp\left(  -\mathrm{i}%
\omega t\right)\label{FourierTrans}\\
&  =\left\vert
{\textstyle\int\limits_{-\infty}^{\infty}}
\mathrm{d}\varepsilon N_{A}(\varepsilon)\exp\left(  -\mathrm{i}\varepsilon
t/\hbar\right)  \right\vert ^{2}.
%\end{align}
\end{eqnarray}
%\end{widetext}
This function measures the probability to find a particle in
the site $A$ at time $t$, provided that the system has had a particle at the
same site at time $t=0$. When the system is isolated, $P_{AA}\left( t\right)
$ oscillates with frequency $\omega _{0}$, which coincides with $\widetilde{%
\omega }$. When the environment is taken into account $P_{AA}\left( t\right)
$ evolves in a more complex way. In spite of this complexity, the evolution
at short times can be described by an exponentially decaying oscillation
with frequency $\widetilde{\omega }$. In the next section we will present a
deeper analysis of all spectral regions and their main dynamical
characteristics.

\section{Parametric Regions}

By analyzing the qualitative features of the LDOS as function of the control
parameter $V_{AB}$ we find five parametric regions separated by well defined
critical values.\ We enumerate them from I to V as they are appearing by
increasing $V_{AB}$ and name them according to the main features in the LDOS
and the behavior of the GF's poles in the complex energy plane. The most
common situation occurs when the two states of the isolated system are mixed
with the environment continuous and hence acquire a finite mean life. This
is region II of \textbf{resolved resonances}, a regime typically described
by the FGR. When the interaction with the environment becomes strong enough,
at the exceptional point appears the nontrivial transition to Region I of
\textbf{collapsed resonances}. This is the regime where exchange narrowing
occurs \cite{Anderson}. In the other extreme, we may consider that the
internal interaction of the system is much stronger than the bandwidth of
the environment's continuous spectrum. Hence, the system's bonding and
anti-bonding states are pushed away from the band according to perturbation
theory and they will remain localized. This is Region V of \textbf{pure
point states}. Region III with \textbf{out-of-band resonant states} and
Region IV of \textbf{virtual states}, would go almost unnoticed unless the
internal and external interactions are very similar, see Fig. \ref{VAB-V0}.
In these cases the separation between system and environment becomes very
delicate.

\begin{figure}[tbp]
\begin{center}
\includegraphics[height=1.7106in,width=3.0727in]
{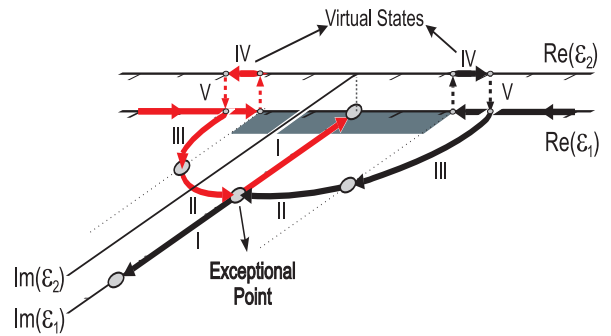}
\end{center}
\caption{(Color online) Paths in the complex energy plane of the two poles
(black and red) of the GF as $V_{AB}$ decreases. Units are arbitrary. They
go from the localized bonding and antibonding states into respective
resonances that eventually collapse at the exceptional point. The bottom
Riemann sheet contains the physical poles, while the upper sheet has the
nonphysical poles. The broader horizontal line in the center represents the
continuous band of environment states. Notice that localized states
transform into virtual states and out-of-band resonances before becoming
well defined resonances.}
\label{RE-Im polos plano}
\end{figure}

\begin{figure}[tbp]
\begin{center}
\includegraphics[height=2.435in,width=3.056in]
{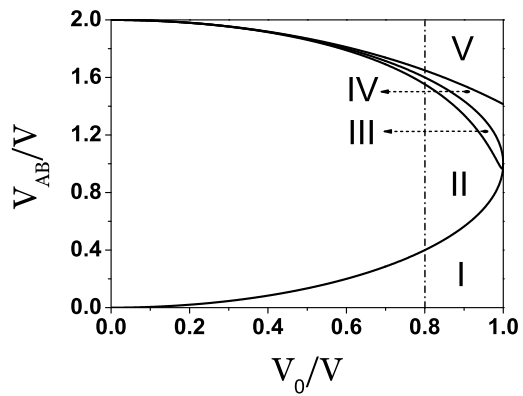}
\end{center}
\caption{Phase diagram of the different regimes as a function of the
dimensionless hoppings $V_{AB}/V$ and $V_{0}/V$. The vertical dash-dotted
line represents the condition used in Figs. \protect\ref{Re + Im-Polos},
\protect\ref{RE-Im polos plano}, \protect\ref{Virtual - Fig}, and \protect\ref%
{LDoS-Figura}.}
\label{VAB-V0}
\end{figure}

The main features of these regions can be identified by following the GF's
poles into the complex plane as represented in Fig. \ref{Re + Im-Polos} and
zoomed in Fig. \ref{Re-Im chico}. We notice that poles in regions III and IV
have \ a real part extending outside the band edges (indicated with an
horizontal dotted line). However, as they enter region IV they loose their
imaginary part. At this point the poles jump up to the second Riemann sheet,
as represented in Fig. \ref{RE-Im polos plano}, and they become virtual
states \cite{Hog95}. This is manifested in the fact they are poles of Eq. %
\ref{Green de inf Sitios} only because the unphysical branch of the
self-energy was used. Hence, they do not show up as peaks or deltas in the
density of states. Indeed, the LDOS given by Eq. \ref{LDoS Overdamped
Equation} integrates to 1 within the band support. Only when the poles reach
the band edge again, see Fig. \ref{RE-Im polos plano}, they return to the
first Riemann sheet to become localized states.

Figure \ref{VAB-V0} displays the dynamical regions in the parameter
space. Notice that if we move along an horizontal line, which
means keeping $V_{AB}$ constant, we are not going to see all the
regions. At least one region would escape the analysis. From
the experimental point of view this implies that, in order to find all the
regimes by controlling only one parameter, it is better to vary the system
frequency than the system environment interaction. For that reason, in the
plots that follow, we only vary the parameter $V_{AB}$ while keeping $V_{0}$
constant at $0.8V$.

In order to ensure consistency, we also studied the behavior of $P_{AA}(t)$
by an exact diagonalization of the Hamiltonian of the finite system. In this
case the environment size is taken large enough so the mesoscopic echoes do
not show up at the times of interest \cite{FP06}. The evaluation of $%
P_{AA}(t)$ allows us to univocally identify the localized states and the
different decay laws of the other regimes. In the next sub-sections we will
show, in more detail, the behavior of the system in each region.
\begin{figure*}[tbp]
\begin{center}
\includegraphics[height=6.0in,width=4.5in]
{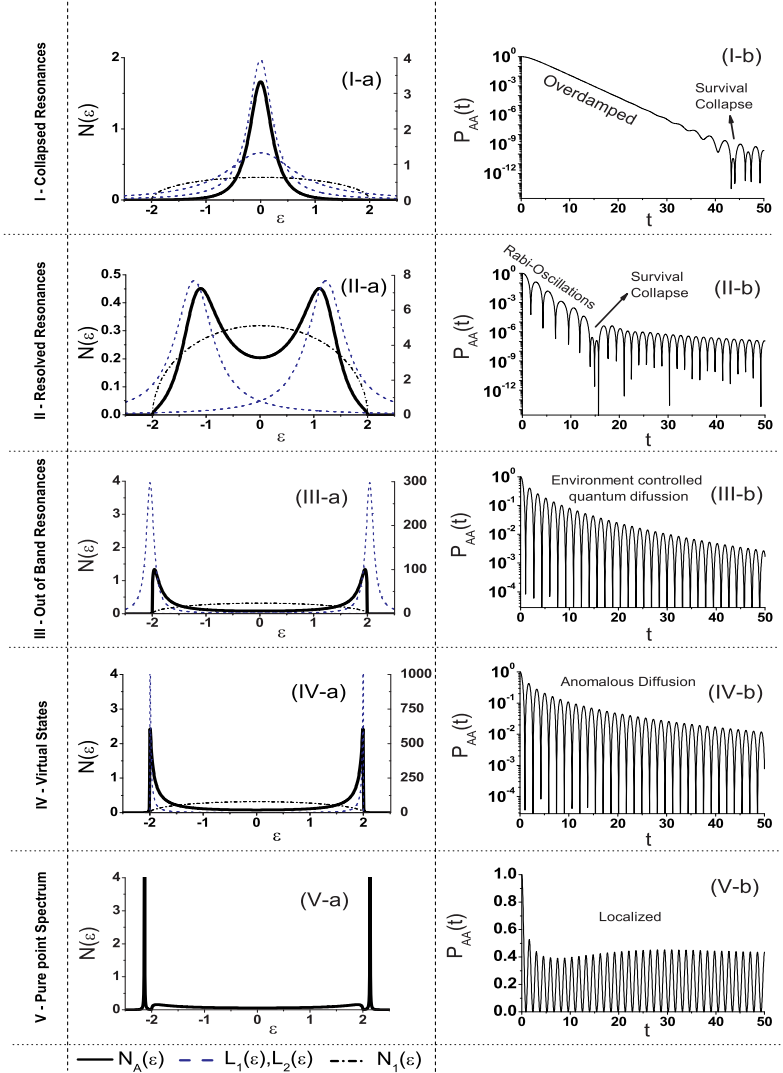}
\end{center}
\caption{(Color online) (a) Left side panels: (Black solid
line) LDOS (in units $1/V)$ for the different regions [ $V_{AB}/V=0.35$~(I),%
$~1.0$(II), $1.58~$(III), $1.62~$(IV), and $1.9~$(V)]. (Blue dotted
line) $L_{1}\left( \protect\varepsilon \right) $ and $L_{2}\left( \protect%
\varepsilon \right) $. (Black dash-dotted line) $N_{1}\left( \protect%
\varepsilon \right) $. The right side scale corresponds to the $L_{1}$ and $%
L_{2}$ plots.(b) Right side panels: Survival probability $%
P_{AA}\left( t\right) $ as a function of time in $\hbar /V$ units.}
\label{LDoS-Figura}
\end{figure*}

\subsection{\label{Overdamp Regime}Region I: Collapsed resonances
(overdamped decay)}

This region is found for:

\begin{equation}
\left\vert V_{AB}\right\vert <\left\vert V-\left( V^{2}-V_{0}^{2}\right)
^{1/2}\right\vert .  \label{Region 1}
\end{equation}

Under this condition, $P_{AA}\left( t\right) $ decays exponentially and
without oscillations until the \textit{survival collapse} time \cite{FP06}.
At this moment the return amplitude from the environment starts to be
comparable with the pure survival amplitude interfering destructively [see
Fig. \ref{LDoS-Figura}(I-b)]. If we set $V_{0}=1V$ we arrive to the case
already treated in Ref. \cite{FP06} where it is analyzed the decay of a
surface spin excitation when it interacts with a spin chain. A similar
dynamics was recently predicted for a quasiparticle excitation at adsorbates
in a surface band \cite{Gumhalter}.

The \textit{real part} of both poles of the\ GF \textit{coincide} with the
site energy, which means that the effective frequency is zero. However,
their respective \textit{imaginary parts} \textit{differ} substantially (see
Fig. \ref{RE-Im polos plano}). One of them moves away from the real axis as
the $\mathcal{S-E}$ interaction increases while the other approaches the real
axis. This means that one state is captured by the environment while the
other becomes isolated by cause\ of the quantum Zeno effect\cite{ADLP06}.

%EndExpansion
In this region $L_{1}$ and $L_{2}$ are two LF centered at $0$, but with
different widths,
\begin{equation}
L_{1,2}\left(  \varepsilon\right)  =C\frac{2\Gamma_{1,2}}{\varepsilon
^{2}+\Gamma_{1,2}^{2}}\text{, }%
\end{equation}
where%
\begin{equation}
C^{2}=\frac{V_{0}^{2}V_{AB}^{2}}{4\Gamma_{1}\Gamma_{2}},%
\end{equation}%
\begin{widetext}
\begin{equation}
\Gamma_{1,2}^{2}=\frac{V_{0}^{4}-V_{AB}^{2}\left(  2V^{2}-V_{0}^{2}\right)
}{2\left(  V^{2}-V_{0}^{2}\right)  }\mp\frac{\sqrt{\left(  V_{0}^{4}%
-V_{AB}^{2}\left(  2V^{2}-V_{0}^{2}\right)  \right)  ^{2}-4V^{2}V_{AB}%
^{4}\left(  V^{2}-V_{0}^{2}\right)  }}{2\left(  V^{2}-V_{0}^{2}\right)  }.
\end{equation}
\end{widetext}
Figure \ref{LDoS-Figura}(I-a) shows the behavior of $N_{A}$, $N_{1}$ and the
Lorentzians functions: $L_{1}$ and $L_{2}$. The centers and the linewidth of
$L_{1}\left(  \varepsilon\right)  $ and $L_{2}\left(  \varepsilon\right)  $
are exactly equal to the real and imaginary part of the poles of the GF, respectively.
\begin{figure}[tbp]
\begin{center}
\includegraphics[height=2.254in,width=3.0652in]
{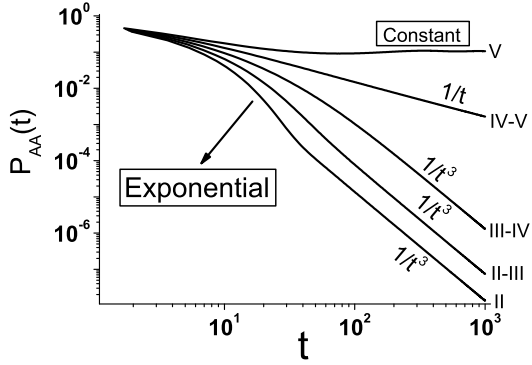}
\end{center}
\caption{Local dynamics in log-log representation for different regimes.
Time is in $\hbar /V$ units. Curves II and V represent evolutions inside
regions II and V, respectively. Curves II-III, III-IV, and IV-V represent
the survival probability at the transitions between regimes.}
\label{LOG-LOG}
\end{figure}

When $V_{AB}$ reaches zero, $\Gamma _{1}$ vanishes, which is consistent with
the fact that the first site becomes completely isolated and the second site
behaves exactly as the case treated in Ref. \cite{FP06}. On the other hand
at the value $V_{AB}=\left\vert V-\left( V^{2}-V_{0}^{2}\right)
^{1/2}\right\vert $ the system presents an EP. In this point both
Lorentzians widths become equal. Beyond this point $\widetilde{\omega }$
starts to grow up (see Fig. \ref{Re + Im-Polos}).\textbf{\ }We can interpret
this critical point as a change of\ representation from an adsorbed atom at
surface of a semi-infinite chain, to an adsorbed dimer that appears when the
initial adatom binds strongly to the surface atom.

\subsection{Region II: Resolved resonances (damped oscillations)}

This regime occurs for values of $V_{AB}$ satisfying%
\begin{widetext}
\begin{equation}
\left\vert V-\left(  V^{2}-V_{0}^{2}\right)  ^{1/2}\right\vert <\left\vert
V_{AB}\right\vert <\left\vert \sqrt{\left(  1+16\frac{V^{2}}{V_{0}^{4}}\left(
V^{2}-V_{0}^{2}\right)  \right)  \left(  \left(  2V^{2}-V_{0}^{2}\right)
-2\sqrt{V^{2}\left(  V^{2}-V_{0}^{2}\right)  }\right)  }\right\vert .
\label{Region 2}%
\end{equation}
\end{widetext}This parametric region is characterized by the
oscillatory-exponential decay of $P_{AA}\left( t\right) $ at short times,
Figs. \ref{LDoS-Figura}(II-b). After the survival collapse it follows a
decay controlled by quantum diffusion ($t^{-3}$). This is better appreciated
in Fig. \ref{LOG-LOG}. Indeed this regime was recently found in a model that
describes the survival probability of an excitation in a multibarrier
superlattice \cite{GarciaCalderon07}. The power law decay is consequence of
the environment's memory effects and corresponds to quantum diffusion from
the \textquotedblleft \textit{edge}\textquotedblright\ of a 1D system.
This is verified by making a Fourier Transform of a LDOS of the form $%
N_{1}(\varepsilon )\propto \varepsilon ^{\nu }\theta \left[ \varepsilon %
\right] $\ which leads to a survival probability with an asymptotic form $%
P(t)\backsim t^{-\left( 2\nu +2\right) }.$ In a \textit{bulk} of $d$
dimensions $\nu =d_{\mathrm{eff}.}/2-1$ with $d_{\mathrm{eff}}=d$. Notably,
our case corresponds to a survival probability of a state in a \textit{%
surface} or \textit{edge, }which decays as a bulk excitation in a higher
effective dimension $d_{\mathrm{eff}}=d+2n$ with $n\leq d$ the order of
the surface. In our case $n=d=1.$

The poles in this region have both real and imaginary parts. At shorts
times, the real part controls the oscillatory behavior [$\widetilde{\omega }%
=2\mathrm{Re}\left( \varepsilon\right) $] and the imaginary part determines
the rate of the exponential decay. On the other hand, at long times the
excitation decays with a $t^{-3}$ law and oscillates with a frequency
determined by the environment band width $4V$.

Here, the critical time $t_{c}$ is the time scale at which the quantum
pathways returning from the environment starts to be comparable to the pure
survival amplitude. Hence, $t_{c}$, which decreases with $V_{AB},$ divides
the exponential decay from the quantum diffusive decay.

The LDOS can be expressed in the same way as Eq. \ref{LDoS Overdamped
Equation}, but in this regime the LFs are centered at symmetric points and
have the same width [see Fig. \ref{LDoS-Figura}(II-a)],%
\begin{equation}
L_{1,2}\left( \varepsilon \right) =C\frac{2\Gamma }{\left( \varepsilon \mp
\varepsilon _{r}\right) ^{2}+\Gamma ^{2}},  \label{L12}
\end{equation}%
with%
\begin{equation}
C^{2}=\frac{V_{0}^{2}V_{AB}^{2}}{4\Gamma ^{2}},
\end{equation}%
\begin{equation}
\Gamma ^{2}=\frac{V_{0}^{4}-V_{AB}^{2}\left( 2V^{2}-V_{0}^{2}\right) }{%
4\left( V^{2}-V_{0}^{2}\right) }+\sqrt{\frac{V^{2}V_{AB}^{4}}{4\left(
V^{2}-V_{0}^{2}\right) }},  \label{Gamma(Er)}
\end{equation}%
\begin{equation}
\varepsilon _{r}^{2}=\frac{V_{AB}^{2}\left( 2V^{2}-V_{0}^{2}\right)
-V_{0}^{4}}{2\left( V^{2}-V_{0}^{2}\right) }+\Gamma ^{2}.  \label{Er}
\end{equation}%
It is interesting to notice that the Rabi frequency could either be slower
than the unperturbed one, when $V_{AB}\ll V,$ or faster, when $V_{AB}\gtrsim
V.~$The precise cross over results from Eq.\ref{Er} when $\varepsilon
_{r}=V_{AB}.$

\subsection{Region III: Out-of-band resonances (environment controlled
quantum diffusion)}

This region is found for $V_{AB}$ satisfying:%
\begin{widetext}
\begin{equation}
\left\vert \sqrt{\left(  1+16\frac{V^{2}}{V_{0}^{4}}\left(  V^{2}-V_{0}%
^{2}\right)  \right)  \left(  \left(  2V^{2}-V_{0}^{2}\right)  -2\sqrt
{V^{2}\left(  V^{2}-V_{0}^{2}\right)  }\right)  }\right\vert <\left\vert
V_{AB}\right\vert <\left\vert V+\left(  V^{2}-V_{0}^{2}\right)  ^{1/2}%
\right\vert . \label{Region III}%
\end{equation}
\end{widetext}
If we only take into account the poles of the GF the system resembles the
Resolved Resonances regime. However, if we compare their LDOS and $%
P_{AA}\left( t\right) $, we conclude that this is a new regime.
\begin{figure}[tbp]
\begin{center}
\includegraphics[height=2.5607in,width=3.0588in]
{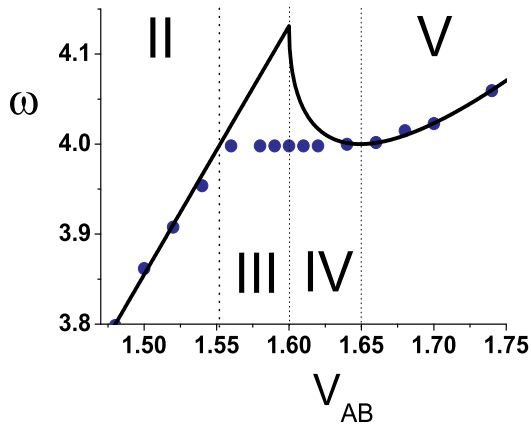}
\end{center}
\caption{(Color online) \textit{Solid line}: frequency $\widetilde{\protect%
\omega }$ evaluated from the poles distance (in units of $V/\hbar $) vs $%
V_{AB}$ (in units of $V)$. \textit{Blue dots}: frequency  fitted from the
dynamics. The value for $V_{0}$ is the same as in Fig. \protect\ref{Re +
Im-Polos}.}
\label{Virtual - Fig}
\end{figure}

In this region, it is no longer possible to distinguish an exponential decay
in $P_{AA}\left( t\right) $ [see Fig. \ref{LDoS-Figura}(III-b)]. This is
because the time $t_{c}$ is too short. This implies that the oscillations
are a direct consequence of the environment. In Fig. \ref{Virtual - Fig} it
is observed that the frequency obtained directly from the numerical
solutions remains constant at $4V$ while $\widetilde{\omega }$ (frequency
evaluated from the poles)\ continues to grow, confirming our analysis.

The expression for the LDOS presented in Eq. \ref{LDoS Overdamped Equation}
is still valid, but the LF are centered outside the band edges [see Fig. \ref%
{LDoS-Figura}(III-a)]. If we analyze the dynamics in term of the LDOS, we
see that the tails of the LF decay as $\varepsilon^{-2}$, which means that
the LDOS will have a Van Hove singularity as $\varepsilon^{1/2}\times
\varepsilon^{-2},$which implies that the local excitation will decay as $%
t^{-3}$ at long times [see Figs. \ref{LDoS-Figura}(III-b) and \ref{LOG-LOG}].

\subsection{Region IV: Virtual states (anomalous diffusion)}

This region corresponds to the range%
\begin{equation}
\left\vert V+\left( V^{2}-V_{0}^{2}\right) ^{1/2}\right\vert <\left\vert
V_{AB}\right\vert <\sqrt{2\left( 2V^{2}-V_{0}^{2}\right) }.  \label{Region 3}
\end{equation}
The poles in this region do not have imaginary part. For this reason it
might be expected that poles were localized states \cite{Economou}. However,
both the LDOS of Fig. \ref{LDoS-Figura}(IV-a) and the dynamics shown in Fig. %
\ref{LDoS-Figura}(IV-b) and Fig. \ref{LOG-LOG} probes that this is not the
case. The reason is that the poles had moved to a second Riemann sheet \cite%
{Hog95, Marcelis} (see Fig. \ref{RE-Im polos plano}). In such case only the
use of the unphysical sign of the self-energy could provide poles. The
dynamics in this regime presents a striking transition between a $t^{-3}$
decay [at $V_{AB}=\left\vert V+\left( V^{2}-V_{0}^{2}\right)
^{1/2}\right\vert $] to a $t^{-1}$ behavior [at\ $V_{AB}=\sqrt{2\left(
2V^{2}-V_{0}^{2}\right) }$]. This is shown in Fig. \ref{LOG-LOG}.

Once again, we can express the LDOS as in Eq. \ref{LDoS Overdamped Equation}%
, however now the interpretation is different. In this region $\varepsilon
_{r}$ is still outside the band edges and yet the value of $\Gamma $ becomes
imaginary. The fact that $\Gamma $ is transformed into an imaginary number
implies that $L_{1}$ and $L_{2}$ are no longer LFs. If we now analyze the
tails of these functions, we observe that they decay as $\varepsilon ^{-1}$.
This leads to a LDOS with a Van Hove singularity of the form $\varepsilon
^{1/2}\times \varepsilon ^{-1}.$Therefore, we can achieve a $t^{-1}$
behavior of $P_{AA}$ at long times. This fact is indeed confirmed by the
observed dynamics (see Fig. \ref{LOG-LOG}). As in the previous subsection, $%
\widetilde{\omega }$ do not follows the observed oscillation frequency [see
Fig. \ref{Virtual - Fig} and \ref{LDoS-Figura}(IV-a)]. Instead, it is fixed
by the environment bandwidth $4V$.

When $V_{AB}$ reaches the value $\sqrt{2\left( 2V^{2}-V_{0}^{2}\right) }$,
there is a change in the nature of the van Hove singularities from $%
\varepsilon^{1/2}$ to $\varepsilon^{-1/2}$. Consequently, $%
P_{AA}(t)$ decays exactly as a $t^{-1}$(see Fig. \ref{LOG-LOG}). From this
point on, the states become localized.

It is interesting to note that while the presence of the virtual states is
not clearly distinguishable in the observable LDOS, the anomalous diffusion,
where $P_{AA}(t)$ moves gradually between $t^{-3}$and $t^{-1}$, should
enable its experimental identification.

\subsection{Region V: Pure point states (localized)}

Finally, the last region appears when,%
\begin{equation}
\left\vert V_{AB}\right\vert >\sqrt{2\left( 2V^{2}-V_{0}^{2}\right) }.
\label{Region 4}
\end{equation}%
Two localized states emerge from the band edges as shown in Fig. \ref%
{LDoS-Figura}(V-a). The poles are real. Figure \ref{Virtual - Fig} shows that $%
\widetilde{\omega }$ recovers its interpretation as the effective system
frequency. In this region the environment renormalization is almost
negligible and its only effect is to slightly correct the value of the
effective frequency. If $V_{AB}$ becomes large enough $\widetilde{\omega }$
reaches $\omega _{0}$. The dynamics in this region, Fig. \ref{LDoS-Figura}(V-b)
is characterized by an oscillatory $P_{AA}(t)$ that only decays for a short
period after which the amplitude of the oscillation remains constant.

\section{Concluding Remarks}

By considering an exactly solvable model for a swap gate in presence of an
environment, we discussed how the memory of the bath affects the dynamics
beyond the Fermi golden rule. The unperturbed Rabi frequency sweeps through
different dynamical and analytic regimes. Depending on the value of the
internal time scales of the system, the environment can induce multiple
phase transitions in the system dynamics.\textbf{\ }Our model shows uncommon
regimes as the exchange narrowing starting at the exceptional point and the
virtual states by moving a single parameter. The fact that all the dynamical
phases appear in the same system, offered the opportunity to study the
transitions between them and hence to determine the precise critical points.

By studying the dynamics, we have characterized the difference between
collapsed resonances, resolved resonances, out-of-band resonances, virtual
states and the pure point states. In particular, we have shown that in the
virtual state regime there is an anomalous quantum diffusive law, which at
long times is observed as a change in the decay law from $t^{-3}$ to $t^{-1}$%
. We have found an expression for the LDOS where the presence of the
resonances becomes explicit. This LDOS is factorized in three terms: a
density of directly connected states (i.e. the LDOS at the environment's
surface) and two Lorentzian Functions whose widths become imaginary in the
virtual state regime. The edges of the LDOS determine the behavior at long
times. It is then clear that the anomalous diffusion is related to the fact
that the van Hove singularities at these edges are modified. This change
occurs when the Lorentzian widths become imaginary.

Note that the complexity of the dynamics for this relatively simple system,
emerges as a consequence of the explicit way in which the environment is
modeled. Details such as out-of-band resonant states and virtual states could
not have been observed in simpler representations of the environment as the\
usual broad-band or the self-consistent Born approximations. The results of
our model system shows that a zero imaginary part of the poles is not enough
as a localization criterion \cite{Economou}. In particular, virtual states
have zero imaginary part, but any local excitation in this parametric regime
shows a complete decay to the environment.

Another parameter that characterizes the dynamics, is the oscillation
frequency. First, in region I (collapsed resonances), before the system
reaches the EP, there is an overdamped decay as expected. When the
unperturbed Rabi frequency $\omega _{0}$ exceeds the critical value, at the
EP, the resonances become resolved (region II).\ $P_{AA}(t)$ show an
exponentially attenuated oscillation with a frequency $\tilde{\omega}$ given
by the poles difference. This frequency $\tilde{\omega}$ goes from 0, at
the EP, to a value higher than $\omega _{0}$. In the region III (out-of-band
resonances) and IV (virtual states) the observed frequency is fully
determined by the environment bandwidth and not by the GF poles, while the
decay follows different power laws. These are particularly stable regimes
for a swap gate since the effective Rabi frequency does not depend on the
internal parameter. Further increase of $\omega _{0}$ leads to Region V of
localized states where the observed frequency is always higher than $\omega
_{0}$ but tends to it as $V_{AB}\rightarrow \infty .$

An interesting message that can be extracted from the phase diagram of Fig. %
\ref{VAB-V0} is that by changing $V_{AB}/V$ one can always move between all
five regimes. However, for small values of the interaction with the
environment $(V_{0}/V)$ the transition to localized states occurs within a
very small range of $V_{AB}/V$ and hence most chances are that it goes
unobserved. The diagram indicates that a clear numerical or experimental
observation of the full dynamical wealth would only be possible for $V_{0}/V$
$\lesssim 1$.

One might wonder how decoherence affects the swap gate. This, of course,
would depend on the specific situation. Consider, for example, a spin chain
with XY interaction as described in Refs. \cite{Danieli-ChPhLett} and \cite%
{ADLP06}. One may assume that the initial condition of the dimer used a swap
gate is in the state $\left\vert 1,0\right\rangle $ while the chain that
acts as environment is in a mixed state with 1/2 occupation at each site. The
environment density will gradually tunnel into the dimer until it becomes a
mixed state itself. In this case, decoherence originates in the mixed
initial condition of the \textquotedblleft bath\textquotedblright\ spins.
While the probability would tends to an asymptotic value, which might differ
from the ergodic 1/2\cite{horacio-mesoscopic}, the Rabi frequency would
remain that evaluated from our coherent calculation. Another situation that
one might conceive is a second environment \ $\mathcal{E}^{\prime }$
yielding a weak decoherent process of strength $\eta $ added homogeneously
throughout the whole system as an imaginary component of the site energies.
In this case, oscillations would simply degrade at a rate $2\eta /\hbar .$
However, a prediction of the effect of strong decoherence requires specific
analysis as it is known that, even at a FGR level, self-consistent
decoherence may lead to dynamical phase transitions \cite%
{DanieliSSC,Horacio-FGR}.

As a final remark, we again mention that because of its simplicity, our
model can be arranged to describe various physical systems beyond spin
chains such as microwave antenna arrays \cite{Sweatlock}, energy shuttling
on plasmon wires made of nanoparticles \cite{Brongersma-Atwater}, arrays of
tunneling coupled optical waveguides \cite{Longui}, periodic elastic arrays
 \cite{Mendez}, or acoustic time reversal cavities \cite{HernanPhysicaB}.
Reciprocally, most of them should present the variate dynamical phenomena
discussed here, provided that one focuses in the proper parameter range.
These examples also suggest possible experimental setups where the
parameters found in this work can be used as a knob enabling to store and
exchange energy.

\begin{acknowledgments}

The authors acknowledge Gonzalo \'{A}lvarez, Hern\'{a}n Calvo, Ernesto
Danieli, and Elena Rufeil-Fiori for sharing their experiences in related
fields. Patricia Levstein and Pablo Serra helped with discussions and useful
references. HMP acknowledges hospitality of MPI-PKS and Abdus Salam ICTP
where discussion with Ingrid Rotter and Maksim Miski-Oglu helped to enrich
this work. The work was made possible through the financial support from
CONICET, ANPCyT, and SeCyT-UNC.
\end{acknowledgments}

\end{document}